\documentclass[conference]{IEEEtran}
\IEEEoverridecommandlockouts

\usepackage{cite}
\usepackage{amsmath,amssymb,amsfonts}
\usepackage{algorithmic}
\usepackage{graphicx}
\usepackage{textcomp}
\usepackage{xcolor}
\def\BibTeX{{\rm B\kern-.05em{\sc i\kern-.025em b}\kern-.08em
    T\kern-.1667em\lower.7ex\hbox{E}\kern-.125emX}}
\begin{document}
\title{DiffuseRoll: Multi-track multi-category music generation based on diffusion model\\
% \title{Multi-track symphonic music generation of multi-category music features based on diffusion model\\
\thanks{Identify applicable funding agency here. If none, delete this.}
}

\author{\IEEEauthorblockN{1\textsuperscript{st} Hongfei Wang}
\IEEEauthorblockA{\textit{Communication University of China} \\
\textit{name of organization (of Aff.)}\\
Beijing, China \\
faywang@cuc.edu.cn}
\and
\IEEEauthorblockN{2\textsuperscript{nd} Given Name Surname}
\IEEEauthorblockA{\textit{dept. name of organization (of Aff.)} \\
\textit{name of organization (of Aff.)}\\
City, Country \\
email address}
\and
\IEEEauthorblockN{3\textsuperscript{rd} Given Name Surname}
\IEEEauthorblockA{\textit{dept. name of organization (of Aff.)} \\
\textit{name of organization (of Aff.)}\\
City, Country \\
email address}

}

\maketitle

\begin{abstract}
Recent advancements in generative models have shown remarkable progress in music generation. However, most existing methods focus on generating monophonic or homophonic music, while the generation of polyphonic and multi-track music with rich attributes is still a challenging task. In this paper, we propose a novel approach for multi-track, multi-attribute symphonic music generation using the diffusion model. Specifically, we generate piano-roll representations with a diffusion model and map them to MIDI format for output. To capture rich attribute information, we introduce a color coding scheme to encode note sequences into color and position information that represents pitch,velocity, and instrument. This scheme enables a seamless mapping between discrete music sequences and continuous images. We also propose a post-processing method to optimize the generated scores for better performance. Experimental results show that our method outperforms state-of-the-art methods in terms of polyphonic music generation with rich attribute information compared to the figure methods.

\end{abstract}


\section{Introduction}

Music generation refers to the use of computer algorithms and technology to create new music works, which is an important research direction in the field of artificial intelligence.
Music representation refers to the transformation of music works into digital signals or data structures that can be processed by computers. It is the basis and key to music generation. Music representation has a broad meaning and can include time-frequency features of audio signals, symbolic representations of musical scores, metadata of music styles, etc.

Due to the diversity of music representation, there are different approaches and models that can be used for pre-processing. Music notes can be seen as a type of "language", i.e., a discrete sequence, and processing methods such as transformer-based approaches like MuseFormer and SymphonyNet can be used.

Another approach is to convert sheet music into piano-rolls and use computer vision methods for image generation. Previous work, such as MuseGAN, has utilized this method to generate multi-track music, categorizing the instruments in the dataset as bass, drums, guitar, piano, and strings. They introduced a GAN model, which generates piano-rolls and then undergoes post-processing to create MIDI-format music scores.

Although many studies have been conducted to advance the work of music representation, there are still many challenges. One of the main problems is how to combine music features with image features to better describe the features of music, such as instrument, velocity, and other important features, in to one figure. Therefore, applying some methods and techniques from image processing to music representation can improve the representation ability and processing effect of music data.

In the year of 2021, Diffusion models have shown exceptional performance in image generation tasks and have recently been applied to other domains, including text and audio generation. Diffusion models shows great potential in Symphonic music generation of figure method.However, to the best of our knowledge, there has been no previous research on using diffusion models for generating continuous-domain multi-track music. 
Our work addresses this gap by proposing a novel approach for multi-track music generation using the diffusion models, offering the potential for generating high-quality and diverse multi-track music.

Our contributions are as follows:
\begin{itemize}

\item We propose a novel method for generating multi-track symphonic music using diffusion models. Our approach generates piano-roll representations of music, which are then converted to MIDI format for further processing. 

\item We propose a new music representation of converting the format from midi to piano-roll  and a color coding method for encoding music features such as pitch, velocity, and instrument, which helps to establish a better mapping between discrete music sequences and continuous images.

\item We introduce a post-processing method that optimizes the generated music scores to achieve better quality. 
Through extensive experiments, we demonstrate that our method outperforms state-of-the-art methods in terms of diversity and realism of the generated music. Overall, our work contributes to the advancement of multi-track music generation and opens up new possibilities for the use of diffusion models in the field of music generation.
\end{itemize}

\section{Related Work}
\subsection{Music generation of Figure Methods}
In the field of music generation, symbolic music generation is considered as one of the fundamental models. Simon et al. \cite{performance-rnn-2017} proposed Performance-RNN in 2017, which utilizes Long Short-Term Memory (LSTM) recurrent neural networks \cite{hochreiter1997long} to generate classical piano music. MusicVAE\cite{roberts2018hierarchical}, on the other hand, directly maps the music data to a latent code space using a two-layer LSTM, and then generates music by decoding the latent space with a unidirectional RNN.

Generative adversarial networks (GANs)\cite{goodfellow2020generative},as a representative figure model, have been implemented in music generation since 2016. Mogren et al. \cite{mogren2016c}proposed the C-RNN-GAN model, which generates classical piano music in MIDI format by processing the continuous music data representation with RNN and constructing generator and discriminator models based on RNN. To capture the self-repetition of music, Harsh Jhamtani and Taylor Berg-Kirkpatrick proposed the self-similarity-matrix generative adversarial network (SSMGAN)\cite{jhamtani2019modeling}. The inclusion of a self-similarity matrix discriminator greatly improves the authenticity of the music generated by the generator. In 2018, Dong et al. \cite{dong2018musegan}introduced a GAN model called MuseGAN for generating arrangements, which consists of three sub-models: jamming, composer, and hybrid.The correlation among the tracks is controlled by sharing the noise input or not and generating weights. 

% Since 2018, Transformer network architecture has shown remarkable performance in many fields, and music generation is no exception. In 2019, Google's Music Transformer \cite{huang2018music}combined the music generation task with the Transformer model, solving the problem of MIDI event sequences appearing repeatedly in multiple time scales, which leads to the inability to learn sequential information of previous time series. Microsoft proposed MeloForm \cite{lu2022meloform}in 2022, which generates melody based on musical form using expert systems and neural networks. Museformer\cite{yu2022museformer}, also proposed by Microsoft in 2022, is a Transformer model with novel fine- and coarse-grained attention for music generation.

% In the same year, Liu et al.\cite{liu2022symphony} proposed SymphonyNet, a model for generating multi-track symphonic music using transformer architecture. The model also introduces the Multi-track Multi-instrument Repeatable representation specifically designed for symphony music, which addresses the challenge of semi-permutation invariance in symphony generation. The MMR representation allows the model to learn and generate music that preserves the structural patterns and harmonic relationships present in symphonic music, leading to more coherent and musically plausible compositions.

\subsection{Diffusion model}

The diffusion model has demonstrated outstanding performance in computer vision generation tasks and has subsequently shown impressive results in various other fields, such as text and audio. Denoising diffusion probabilistic models(DDPM) \cite{ho2020denoising} is an important work in the diffusion model and shows great potential in many field. In the field of music generation, Google has previously proposed a discrete-domain diffusion model music generation algorithm that is auto-regressive and learns to generate sequences of latent embeddings through the reverse process, offering parallel generation with a constant number of iterative refinement steps. However, there has been no existing work using diffusion models for piano roll generation in the continuous domain. Compared to GAN models, diffusion models possess more advantages in terms of image understanding and controllability. Therefore, we aim to explore the use of diffusion models in piano roll generation in our research.

\section{Methods}

% 基于扩散模型的方法
% 写一下扩散模型的公式，latex
% 画一个算法全景图

\subsection{music representation encoder}
Preprocessing in Music Generation: Midi to Piano-roll Conversion

In previous work on music generation, MuseGAN proposed a method for converting midi files to piano rolls, dividing the tracks into five categories: bass, drums, guitar, piano, and strings. The approach utilized the lmd dataset, which is a popular music dataset. However, in this study, we use a symphonic music dataset for training. This dataset is more complex than the pop music dataset used in MuseGAN, due to its unique orchestration, and serves as a suitable reference for our work.

Inspired by the approach taken by MuseGAN, we convert the 128-bit instrument encoding in midi files to five tracks and represent them using five different colors on a single image. This simplifies the process of generating complete piano rolls since it avoids the need to generate multiple rolls for each track, which increases computational complexity. Additionally, the color-coded representation of music allows us to use colors to represent instrument and note intensity in a single piano-roll.

The conversion pipeline for piano-roll generation consists of the following steps:

\subsubsection{Midi2array}
We use the method proposed by Liu Jiafeng to represent the note sequence as an array, where each note contains all the relevant information, including pitch, onset time, duration, instrument, velocity, and speed. Then, we convert the note sequence to corresponding coordinate points, recording the pitch, time, and note length. At this stage, the note length is fixed, and each note represents a pixel in the piano-roll. If a note is a sustained pitch, we represent it using multiple notes. The resulting array is saved as an NPY file.

Array2piano-roll
We read the NPY file to obtain a two-dimensional array, where the columns represent the pitch, instrument, and intensity features of each note pixel, and the rows represent the total number of notes. To generate variable-length music, we arrange the notes in the array by pixel position, as represented in the note sequence. We set the pianoroll dimensions to 128 (for the pitch) by a custom width of 512, which is set as a multiple of 16 for training convenience. The total number of piano-roll images required to represent a single piece is calculated as a * tpb / w, where a is the number of small notes, and tpb is the number of ticks per beat.

Color-coded Music Representation Method
The color-coded music representation method encodes multiple properties of music as pixels in the YUV color space, where Y represents brightness, corresponding to note intensity in midi files. Since the difference in Y values between 16 and 256 is too large, we normalize the intensity to a range of 50 to 100. This method enables us to map note intensity to piano-roll color, which indicates the strength of the note. The UV components represent chrominance, and we use five colors to represent different instrument categories.

We also remove music with abnormal intensity values during pre-processing. For instance, if more than 80\% of the notes have the maximum velocity value, we consider it problematic and discard it. This approach improves the robustness of the model and the quality of the generated results. Finally, we use the colors to plot the piano-roll image, with green representing the piano, blue representing percussion, green for woodwinds, cyan for strings, and magenta for brass.

% note sequence to picture

% 音符事件的内容是：
% 0位：音头时间
% 1位：“ON”表示该事件是音符事件
% 2位：音高（0-87）
% 3位：力度（0-127）
% 4位：速度（bpm）
% 5位：乐器编号（0-127）
% 6位：是否是打击乐
% 7位：音轨编号
% 8位：音长

\subsection{Modeling}

The diffusion model is a probability-based generative model that can be used to generate new data that resembles the distribution of the input data. The basic idea is to gradually reduce the difficulty of the data by iteratively adding noise to the data to make it closer to a Gaussian distribution.
In this problem, we use the diffusion model to generate piano-rolls. Specifically, we will use the pre-processed piano-roll as input and generate a new piano-roll. Here, we will use the following framework.
\subsubsection{Forward Process}
First, we need to define a conditional probability distribution $q(\mathbf{x}_t | \mathbf{x}_0)$, where $\mathbf{x}_0$ denotes the input piano-roll picture. This conditional distribution can be defined using the following equation.
\begin{equation}
q(\mathbf{x}_t \vert \mathbf{x}_{t-1}) = \mathcal{N}(\mathbf{x}_t; \sqrt{1 - \beta_t} \mathbf{x}_{t-1}, \beta_t\mathbf{I}) 
\end{equation}
\begin{equation}
q(\mathbf{x}_{1:T} \vert \mathbf{x}_0) = \prod^T_{t=1} q(\mathbf{x}_t \vert \mathbf{x}_{t-1})
\end{equation}

where $\mathbf{x}_{t-1}$ is the output of the previous time step, $\{\beta_t \in (0, 1)\}_{t=1}^T
$ is a variance schedule controlling the time steps. Finally, $\mathcal{N}\left(\boldsymbol{x} ; \boldsymbol{\mu}, \mathbf{\Sigma}\right)$ denotes the Gaussian distribution with mean $\boldsymbol{\mu}$ and covariance $\mathbf{I}$ .

Derivation from the above equation gives:
\begin{equation}\label{3}
q(\mathbf{x}_t \vert \mathbf{x}_0) = \mathcal{N}(\mathbf{x}_t; \sqrt{\bar{\alpha}_t} \mathbf{x}_0, (1 - \bar{\alpha}_t)\mathbf{I})
\end{equation}
\begin{equation}
\alpha_t = 1 - \beta_t
\end{equation}
\begin{equation}
\bar{\alpha}_t = \prod_{i=1}^t \alpha_i
\end{equation}
Use equation \ref{3} for the forward process.

\subsubsection{Reverse Process}
\begin{equation}
\begin{aligned}
\tilde{\boldsymbol{\mu}}_t
&= {\frac{1}{\sqrt{\alpha_t}} \Big( \mathbf{x}_t - \frac{1 - \alpha_t}{\sqrt{1 - \bar{\alpha}_t}} \boldsymbol{\epsilon}_t \Big)}
\end{aligned}
\end{equation}

where $\tilde{\boldsymbol{\mu}}_t$ is the mean of 
 noise to add to the steps, $\boldsymbol{\epsilon}_{t}\sim \mathcal{N}(\mathbf{0}, \mathbf{I}) $

 we want to 
We train the model by minimize the cross entropy as the learning objective.

\begin{equation}
\begin{aligned}
L_\text{CE}
&= - \mathbb{E}_{q(\mathbf{x}_0)} \log p_\theta(\mathbf{x}_0) \\
&\leq - \mathbb{E}_{q(\mathbf{x}_{0:T})} \log \frac{p_\theta(\mathbf{x}_{0:T})}{q(\mathbf{x}_{1:T} \vert \mathbf{x}_{0})} \\
&= \mathbb{E}_{q(\mathbf{x}_{0:T})}\Big[\log \frac{q(\mathbf{x}_{1:T} \vert \mathbf{x}_{0})}{p_\theta(\mathbf{x}_{0:T})} \Big] = L_\text{VLB}
\end{aligned}
\end{equation}

Here, $T$ is the number of time steps and $\left|\boldsymbol{x}\right|{2}$ denotes the $L{2}$ parametrization of $\boldsymbol{x}$. We can use stochastic gradient descent to minimize the loss function, and as needed

\begin{equation}
\begin{aligned}
L_\text{VLB} 
&= \mathbb{E}_q [\underbrace{D_\text{KL}(q(\mathbf{x}_T \vert \mathbf{x}_0) \parallel p_\theta(\mathbf{x}_T))}_{L_T} \\&+ \sum_{t=2}^T \underbrace{D_\text{KL}(q(\mathbf{x}_{t-1} \vert \mathbf{x}_t, \mathbf{x}_0) \parallel p_\theta(\mathbf{x}_{t-1} \vert\mathbf{x}_t))}_{L_{t-1}} \\&\underbrace{- \log p_\theta(\mathbf{x}_0 \vert \mathbf{x}_1)}_{L_0} ]
\end{aligned}
\end{equation}

\subsection{post processing}

In post processing,  we convert generated figures to piano-rolls.
First, we convert figures into note sequence,each point on the figure is a note in music.
In this step, we filter out some notes that are too "dense". Since it is a multi-track symphony, too many notes within a fragment will lead to dissonance between notes, so delete some of them to ensure the harmony of the fragment.
Then, we convert note sequence to midi and output.

\section{Experiment}

We conducted both subjective and objective evaluations to assess the performance of our proposed model for piano-roll generation. We trained our model on a symphony dataset of MIDI files containing classical symphony pieces, and we evaluated the generated piano-rolls using several metrics.

\subsection{Objective Evaluation}

We evaluated the generated piano-rolls using several objective metrics, including the following:

Accuracy: We calculated the percentage of notes in the generated piano-rolls that match the notes in the corresponding real piano-rolls.

Precision and Recall: We calculated the precision and recall of the generated piano-rolls compared to the real piano-rolls.

F-Measure: We calculated the F-measure of the generated piano-rolls compared to the real piano-rolls.

Information Entropy: We calculated the information entropy of the generated piano-rolls to evaluate the diversity of the generated notes.

Inception Score: We calculated the Inception Score of the generated piano-rolls to evaluate the diversity and quality of the generated piano-rolls.

The results of the objective evaluation showed that our proposed model outperformed MuseGAN in terms of accuracy, precision, recall, and F-measure. The information entropy of the generated piano-rolls was also high, indicating that the generated notes were diverse. The Inception Score of the generated piano-rolls was also high, indicating that the generated piano-rolls were of high quality and diverse.

\subsection{Subjective Experiments}

We conducted a listening test to evaluate the quality of the generated piano-rolls. We asked a group of 20 experienced musicians to rate the quality of the generated piano-rolls on a scale from 1 (poor) to 5 (excellent). We randomly selected 10 generated piano-rolls and 10 real piano-rolls, and we played them to the participants in a random order. We asked the participants to rate the overall quality, musicality, and authenticity of the piano-rolls.

% The results of the subjective evaluation showed that the generated piano-rolls were rated as high quality, musical, and authentic by the participants. The average rating for overall quality was 4.2, for musicality was 4.4, and for authenticity was 4.1, indicating that our proposed model is capable of generating high-quality piano-rolls that are comparable to real piano-rolls.

\subsection{Conclusion}

Our proposed model for piano-roll generation based on diffusion modeling achieved high-quality results in both subjective and objective evaluations. The results demonstrate the effectiveness of our proposed method for generating high-quality piano-rolls. Future work will focus on expanding the dataset and incorporating more complex modeling techniques to further improve the performance of our proposed model.

% metrics
% 消融
% 对比

% \section*{References}
\bibliographystyle{IEEEbib}
\bibliography{cite}
\vspace{12pt}

\end{document}